\newif\ifdraft\draftfalse
\newif\ifanon\anonfalse

\documentclass{svproc}
\usepackage{algorithm}
\usepackage{algpseudocode}
\usepackage{amsmath}
\usepackage{amsfonts}
\usepackage{color}
\usepackage{graphicx}
\usepackage{multirow}
\usepackage{url}
\usepackage{cite}

\newcommand{\ottdrule}[4][]{{\displaystyle\frac{\begin{array}{l}#2\end{array}}{#3}\quad\ottdrulename{#4}}}

\newcommand{\ottpremise}[1]{ #1 \\}
\newenvironment{ottdefnblock}[3][]{ \framebox{\mbox{#2}} \quad #3 \\[0pt]}{}

\newcommand{\ottnt}[1]{\mathit{#1}}
\newcommand{\ottmv}[1]{\mathit{#1}}

\newcommand{\ottsym}[1]{#1}

\newcommand{\ottdrulename}[1]{\textsc{#1}}

\newcommand{\ottdruleTXXHole}[1]{\ottdrule[#1]{%
}{
\Gamma  \vdash  \left[ \, \right] \,  \mathrel{:}  \, \ottnt{T}}{%
{\ottdrulename{T\_Hole}}{}%
}}

\newcommand{\ottdruleTXXVar}[1]{\ottdrule[#1]{%
\ottpremise{\mathit{x}  \mathord{:}  \ottnt{T} \,  \in  \, \Gamma}%
}{
\Gamma  \vdash  \mathit{x} \,  \mathrel{:}  \, \ottnt{T}}{%
{\ottdrulename{T\_Var}}{}%
}}

\newcommand{\ottdruleTXXAbs}[1]{\ottdrule[#1]{%
\ottpremise{\Gamma  \ottsym{,}  \mathit{x}  \mathord{:}  \ottnt{S}  \vdash  \ottnt{M} \,  \mathrel{:}  \, \ottnt{T}}%
}{
\Gamma  \vdash  \lambda  \mathit{x}  \ottsym{.}  \ottnt{M} \,  \mathrel{:}  \, \ottnt{S} \,  \mathord{\rightarrow}  \, \ottnt{T}}{%
{\ottdrulename{T\_Abs}}{}%
}}

\newcommand{\ottdruleTXXApp}[1]{\ottdrule[#1]{%
\ottpremise{ \Gamma  \vdash  \ottnt{M} \,  \mathrel{:}  \, \ottnt{S} \,  \mathord{\rightarrow}  \, \ottnt{T}  \quad  \Gamma  \vdash  \ottnt{N} \,  \mathrel{:}  \, \ottnt{S} }%
}{
\Gamma  \vdash  \ottnt{M} \, \ottnt{N} \,  \mathrel{:}  \, \ottnt{T}}{%
{\ottdrulename{T\_App}}{}%
}}

\newcommand{\ottdruleTXXPair}[1]{\ottdrule[#1]{%
\ottpremise{ \Gamma  \vdash  \ottnt{M} \,  \mathrel{:}  \, \ottnt{S}  \quad  \Gamma  \vdash  \ottnt{N} \,  \mathrel{:}  \, \ottnt{T} }%
}{
\Gamma  \vdash  \ottsym{(}  \ottnt{M}  \ottsym{,}  \ottnt{N}  \ottsym{)} \,  \mathrel{:}  \, \ottnt{S} \,  \mathord{\times}  \, \ottnt{T}}{%
{\ottdrulename{T\_Pair}}{}%
}}

\newcommand{\ottdruleTXXCasePair}[1]{\ottdrule[#1]{%
\ottpremise{ \Gamma  \vdash  \ottnt{M} \,  \mathrel{:}  \, \ottnt{S} \,  \mathord{\times}  \, \ottnt{T}  \quad  \Gamma  \ottsym{,}  \mathit{x}  \mathord{:}  \ottnt{S}  \ottsym{,}  \mathit{y}  \mathord{:}  \ottnt{T}  \vdash  \ottnt{N} \,  \mathrel{:}  \, \ottnt{T'} }%
}{
\Gamma  \vdash  \mathsf{case} \, \ottnt{M} \, \mathsf{of} \, \ottsym{(}  \mathit{x}  \ottsym{,}  \mathit{y}  \ottsym{)}  \rightarrow  \ottnt{N} \,  \mathrel{:}  \, \ottnt{T'}}{%
{\ottdrulename{T\_CasePair}}{}%
}}

\newcommand{\ottdruleTXXLeft}[1]{\ottdrule[#1]{%
\ottpremise{\Gamma  \vdash  \ottnt{M} \,  \mathrel{:}  \, \ottnt{S}}%
}{
\Gamma  \vdash  \mathsf{Left} \, \ottnt{M} \,  \mathrel{:}  \, \ottnt{S} \,  \mathord{+}  \, \ottnt{T}}{%
{\ottdrulename{T\_Left}}{}%
}}

\newcommand{\ottdruleTXXRight}[1]{\ottdrule[#1]{%
\ottpremise{\Gamma  \vdash  \ottnt{M} \,  \mathrel{:}  \, \ottnt{T}}%
}{
\Gamma  \vdash  \mathsf{Right} \, \ottnt{M} \,  \mathrel{:}  \, \ottnt{S} \,  \mathord{+}  \, \ottnt{T}}{%
{\ottdrulename{T\_Right}}{}%
}}

\newcommand{\ottdruleTXXCaseSum}[1]{\ottdrule[#1]{%
\ottpremise{ \Gamma  \vdash  \ottnt{M} \,  \mathrel{:}  \, \ottnt{S} \,  \mathord{+}  \, \ottnt{T}  \quad   \Gamma  \ottsym{,}  \mathit{x}  \mathord{:}  \ottnt{S}  \vdash  \ottnt{N_{{\mathrm{1}}}} \,  \mathrel{:}  \, \ottnt{T'}  \quad  \Gamma  \ottsym{,}  \mathit{y}  \mathord{:}  \ottnt{T}  \vdash  \ottnt{N_{{\mathrm{2}}}} \,  \mathrel{:}  \, \ottnt{T'}  }%
}{
\Gamma  \vdash  \mathsf{case} \, \ottnt{M} \, \mathsf{of} \, \ottsym{\{} \, \mathsf{Left} \, \mathit{x}  \rightarrow  \ottnt{N_{{\mathrm{1}}}}  \ottsym{;} \, \mathsf{Right} \, \mathit{y}  \rightarrow  \ottnt{N_{{\mathrm{2}}}}  \ottsym{\}} \,  \mathrel{:}  \, \ottnt{T'}}{%
{\ottdrulename{T\_CaseSum}}{}%
}}

\usepackage{stmaryrd}

\newcommand\sect[1]{Section~\ref{sec:#1}}
\newcommand\fig[1]{Figure~\ref{fig:#1}}
\newcommand\tbl[1]{Table~\ref{tbl:#1}}

\newcommand{\Rule}[2]{\ensuremath{\text{{\sc{{#1}-{#2}}}}}}
\newcommand{\T}{\Rule{T}}

\newcommand{\seqseq}{\textsc{seq2seq}}

\newcommand\NextTerms{\textsc{GenCandidates}}

\newcommand\COSTBASE[3]{\ensuremath{#1_{#2}(#3)}}
\newcommand\COSTNAME{\textit{cost}}
\newcommand\COST{\COSTBASE{\COSTNAME}}
\newcommand\COSTed{\COSTBASE{\COSTNAME^\textit{{\tiny\,}ed}}}
\newcommand\COSTim{\COSTBASE{\COSTNAME^\textit{{\tiny\,}im}}}
\newcommand\COSTbf{\COSTBASE{\COSTNAME^\textit{{\tiny\,}bf}}}

\newcommand{\SYNTHESIS}{\textsc{Synthesis}}
\newcommand{\SYNTHESISed}{\ensuremath{\SYNTHESIS^\textit{ed}}}
\newcommand{\SYNTHESISim}{\ensuremath{\SYNTHESIS^\textit{im}}}
\newcommand{\SYNTHESISbf}{\ensuremath{\SYNTHESIS^\textit{bf}}}

\newfloat{procedure}{htbp}{loa}
\floatname{procedure}{Procedure}

\algnewcommand\algorithmicforeach{\textbf{for each}}
\algdef{S}[FOR]{ForEach}[1]{\algorithmicforeach\ #1\ \algorithmicdo}

\begin{document}
\mainmatter              
\title{Towards Proof Synthesis\\ Guided by Neural Machine Translation\\ for Intuitionistic Propositional Logic}
\titlerunning{Towards Proof Synthesis Guided by Neural Machine Translation for IPL}
\ifanon
\author{Anon.}
\else
\author{Taro Sekiyama$^1$ \and Akifumi Imanishi$^2$ \and Kohei Suenaga$^{2,3}$}
\authorrunning{Taro Sekiyama, Akifumi Imanishi, and Kohei Suenaga}
\institute{IBM Research -- Tokyo \and Kyoto University \and JST PRESTO}
\fi
%

\maketitle              

\begin{abstract}
  Inspired by the recent evolution of \emph{deep neural networks} (DNNs) in
  machine learning, we explore their application to PL-related topics.  This
  paper is the first step towards this goal; we propose a proof-synthesis method for
  the negation-free propositional logic in which we use a DNN to obtain a guide
  of proof search.  The idea is to view the proof-synthesis problem as a
  \emph{translation from a proposition to its proof}.  We train seq2seq, which
  is a popular network in neural machine translation, so that it generates a proof
  encoded as a $\lambda$-term of a given proposition.  We implement the whole
  framework and empirically observe that a generated proof term is close to a
  correct proof in terms of the tree edit distance of AST.  This observation
  justifies using the output from a trained seq2seq model as a guide for proof
  search.
\end{abstract}

\section{Introduction}
\label{sec:intro}

\emph{Deep neural networks} (DNNs) saw a great success and have become
one of the most popular technologies in machine learning.  They are
especially good at solving problems in which one needs to discover
certain patterns in problem instances (e.g., image
classification~\cite{He_2016_CVPR,DBLP:journals/corr/SimonyanZ14a,DBLP:conf/cvpr/SzegedyLJSRAEVR15},
image generation~\cite{DBLP:conf/icml/GregorDGRW15,NIPS2015_5773}, and
speech
recognition~\cite{DBLP:journals/taslp/DahlYDA12,DBLP:journals/corr/abs-1303-5778}).

Compared to the huge success in these problems, their application to
PL-related problems such as program synthesis and automated theorem
proving is, in spite of recent
progress~\cite{DBLP:journals/corr/BhatiaS16,DBLP:journals/corr/ZarembaS14,DBLP:journals/corr/DevlinUBSMK17,DBLP:journals/corr/BalogGBNT16,DBLP:conf/aaai/MouLZWJ16},
yet to be fully explored.  This is partly because the following gap
between the PL-related areas and the areas where DNNs are competent:
\begin{itemize}
\item The output of a DNN is not guaranteed to be correct; its
  performance is often measured by the ratio of the correct responses
  with respect to the set of test data.  However, at least in the
  traditional formulation of PL-related problems, the answer is
  required to be fully correct.
\item It is nontrivial how to encode an instance of a PL-related
  problem as an input to a DNN.  For example, program synthesis is a
  problem of generating a program $P$ from its specification $S$.
  Although typical representations of $P$ and $S$ are abstract syntax
  trees, feeding a tree to a DNN requires nontrivial
  encoding~\cite{DBLP:conf/aaai/MouLZWJ16}.
\end{itemize}

This paper reports our preliminary experimental result of the
application of DNNs to a PL-related problem, \emph{proof
  synthesis of the intuitionistic propositional logic}.
Proof synthesis leads to solve problems of automated theorem proving, which is one
of the most fundamental problem in the theory of PL and has long
history in computer science.  Automated theorem proving is
also an important tool for program verification, where the correctness
of a program is reduced to the validity of verification conditions.
It is also of interest for \emph{program synthesis} because automated
proof synthesis can be seen as an automated program synthesis via
the Curry--Howard isomorphism~\cite{Sorensen:2006:LCI:1197021}.



Concretely, we propose a procedure that solves the following problem:
\begin{description} \item[\textbf{Input}] A proposition $\ottnt{T}$ of the
propositional logic represented as an AST;\footnote{Currently, we train
  and test the model with propositions of the negation-free fragment
  of this logic.} \item[\textbf{Output}] A proof term $\ottnt{M}$ of $\ottnt{T}$
represented as an AST of the simply typed $\lambda$-calculus extended with pairs
and sums. 
\end{description} 
One of the main purposes of the present work is to measure the
baseline of the proof synthesis with DNNs.
The present paper confirms how a ``vanilla'' DNN framework is smart for our
problem.  As we describe below, we observed that such an untuned DNN indeed works
quite well.
%

In order to apply DNNs to our problem as easily as possible, we take
the following (admittedly simple-minded) view: proof synthesis can be
viewed as \emph{translation from a proposition to a proof term}.
Therefore, we should be able to apply \emph{neural machine translation},
machine translation that uses a deep neural network inside, to the
proof-synthesis problem, just by expressing both a proposition and a
proof as sequences of tokens.


We adopt a \emph{sequence-to-sequence ({\seqseq})
  model}~\cite{DBLP:journals/corr/ChoMGBSB14,DBLP:journals/corr/SutskeverVL14,DBLP:journals/corr/BahdanauCB14},
which achieves good performance in
English--French translation~\cite{DBLP:journals/corr/SutskeverVL14}, for the
proposition--proof translation and train it on a set of proposition--proof
pairs.
Although the trained model generates correct proofs for many
propositions (see \tbl{eval-model} in \sect{exp}; the best model generates
correct proofs for almost half of the benchmark problems),
it sometimes generates (1) a grammatically incorrect token sequence or
(2) an ill-typed response.  As a remedy to these incorrect responses, our
procedure postprocesses the response to obtain a correct proof term.



\begin{figure}[t] \centering
 \includegraphics[width=.95\textwidth,clip]{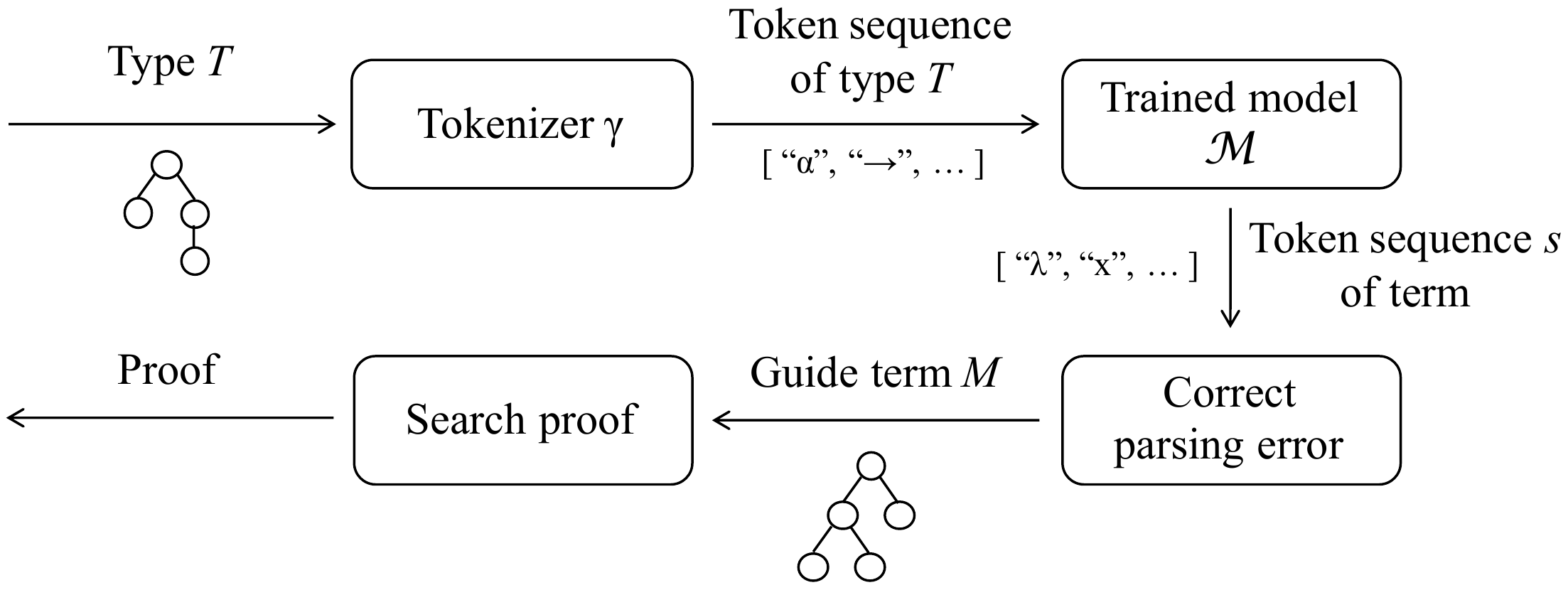}
 \caption{An overview of the proof-synthesis system.}
 \label{fig:system}
\end{figure}

\fig{system} overviews our proof-synthesis procedure.  We explain the important
components:
\begin{itemize}
\item The core of our proof-synthesis method is the neural network $\mathcal{M}$, which
  takes the token-sequence representation $\gamma \, \ottsym{(}  \ottnt{T}  \ottsym{)}$ of a
  proposition $\ottnt{T}$ as an input.  $\mathcal{M}$ is trained to generate a
  proof term of the given proposition; therefore, the output $\mathcal{S}$ of
  $\mathcal{M}$ from $\gamma \, \ottsym{(}  \ottnt{T}  \ottsym{)}$ is expected to be ``close'' to the
  token-sequence representation of a correct proof term of $\ottnt{T}$.
\item The generated token sequence $\mathcal{S}$ may be grammatically incorrect.
  In order to compute a parse tree $\ottnt{M}$ from such an incorrect sequence, we
  apply Myers' algorithm~\cite{DBLP:journals/ipl/Myers95} that
  produces a grammatically correct token sequence that is closest
  to $\mathcal{S}$ in terms of the edit distance.
\item Using the obtained parse tree $\ottnt{M}$ as a guide, our procedure
  searches for a correct proof of $\ottnt{T}$.  To this end, we enumerate parse
  trees in the ascending order of the tree edit distance proposed by
  Zhang et al.~\cite{DBLP:journals/siamcomp/ZhangS89}.  In order to
  check whether an enumerated tree $\ottnt{M'}$ is a correct proof term of
  $\ottnt{T}$, we pass it to a proof checker.  In the current implementation,
  we translate $\ottnt{M'}$ to a Haskell program and typecheck it using
  Haskell interpreter GHCi.  If it typechecks and has $\ottnt{T}$, then, via the
  Curry--Howard isomorphism, we can conclude that $\ottnt{M'}$ is a correct proof
  term of $\ottnt{T}$.
\end{itemize}

We remark that our proof-synthesis procedure is not complete.  Indeed, it does
not terminate if a proposition that is not an intuitionistic tautology is
passed.  We do not claim that we propose the best proof-synthesis procedure,
because a sound and complete proof-synthesis algorithm is known in the
intuitionistic logic~\cite{Sorensen:2006:LCI:1197021}.  The purpose of the
present work is rather exploratory; we show the possibility of DNNs, especially
neural machine translation, for the problem of automated theorem proving.



The rest of the paper is organized as follows.  \sect{language} defines the
target logic as a variant of the simply typed $\lambda$-calculus; \sect{nn}
explains the sequence-to-sequence neural network which we use for proof
synthesis; \sect{synthesis} presents the proof-synthesis procedure; \sect{exp}
describes the experiments; \sect{relatedwork} discusses related work;
and \sect{conclusion} concludes.

\section{Language}
\label{sec:language}

This section fixes the syntax for propositions and proof terms.  Based
on the Curry--Howard isomorphism, we use the notation of the simply typed
$\lambda$-calculus extended with pairs and sums.  We hereafter
identify a type with a proposition and a $\lambda$-term with a proof.


\begin{figure}[t]
 \[\begin{array}{lll}
  \ottnt{S}, \ottnt{T} &::=& \alpha \mid \ottnt{S} \,  \mathord{\rightarrow}  \, \ottnt{T} \mid \ottnt{S} \,  \mathord{\times}  \, \ottnt{T} \mid \ottnt{S} \,  \mathord{+}  \, \ottnt{T} \\
  \ottnt{M}, \ottnt{N} &::=&  \left[ \, \right]  \mid \mathit{x} \mid \lambda  \mathit{x}  \ottsym{.}  \ottnt{M} \mid \ottnt{M} \, \ottnt{N} \mid
                     \ottsym{(}  \ottnt{M}  \ottsym{,}  \ottnt{N}  \ottsym{)} \mid \mathsf{case} \, \ottnt{M} \, \mathsf{of} \, \ottsym{(}  \mathit{x}  \ottsym{,}  \mathit{y}  \ottsym{)}  \rightarrow  \ottnt{N} \mid \\ &&
                     \mathsf{Left} \, \ottnt{M} \mid \mathsf{Right} \, \ottnt{M} \mid
                     \mathsf{case} \, \ottnt{M} \, \mathsf{of} \, \ottsym{\{} \, \mathsf{Left} \, \mathit{x}  \rightarrow  \ottnt{N_{{\mathrm{1}}}}  \ottsym{;} \, \mathsf{Right} \, \mathit{y}  \rightarrow  \ottnt{N_{{\mathrm{2}}}}  \ottsym{\}} \\
  \Gamma        &::=&  \emptyset  \mid \Gamma  \ottsym{,}  \mathit{x}  \mathord{:}  \ottnt{T}
 \end{array}\]

 \caption{Syntax.}
 \label{fig:syntax}
\end{figure}

\begin{figure}[t]

 \framebox{$\Gamma  \vdash  \ottnt{M} \,  \mathrel{:}  \, \ottnt{T}$} \quad \textbf{Typing}
 \begin{center}
  $\ottdruleTXXHole{}$ \hfil
  $\ottdruleTXXVar{}$ \hfil
  $\ottdruleTXXAbs{}$ \\[3ex]
  $\ottdruleTXXApp{}$ \hfil
  $\ottdruleTXXPair{}$ \\[3ex]
  $\ottdruleTXXCasePair{}$ \\[3ex]
  $\ottdruleTXXLeft{}$ \hfil
  $\ottdruleTXXRight{}$ \\[3ex]
  $\ottdruleTXXCaseSum{}$ \hfil
 \end{center}
 \caption{Typing rules.}
 \label{fig:typing}
\end{figure}

\fig{syntax} shows the syntax of the target language.
We use metavariables $\mathit{x},\mathit{y},\mathit{z},\dots$ for variables.
The target language is an extension of the simply typed $\lambda$-calculus with
products, sums, and \emph{holes}.
%
We use a hole in the synthesis procedure described later to represent
a partially synthesized term.
%
Since the syntax is standard (except holes), we omit an
explanation of each construct.  We also omit the dynamic semantics of
the terms; it is not of interest in the present paper.
Free and bound variables are defined in the standard way: a lambda
abstraction $\lambda  \mathit{x}  \ottsym{.}  \ottnt{M}$ binds $\mathit{x}$ in $\ottnt{M}$; a case expression for
pairs $\mathsf{case} \, \ottnt{M} \, \mathsf{of} \, \ottsym{(}  \mathit{x}  \ottsym{,}  \mathit{y}  \ottsym{)}  \rightarrow  \ottnt{N}$ binds $\mathit{x}$ and $\mathit{y}$ in $\ottnt{N}$;
a case expression for sums $\mathsf{case} \, \ottnt{M} \, \mathsf{of} \, \ottsym{\{} \, \mathsf{Left} \, \mathit{x}  \rightarrow  \ottnt{N_{{\mathrm{1}}}}  \ottsym{;} \, \mathsf{Right} \, \mathit{y}  \rightarrow  \ottnt{N_{{\mathrm{2}}}}  \ottsym{\}}$
binds $\mathit{x}$ and $\mathit{y}$ in $\ottnt{N_{{\mathrm{1}}}}$ and $\ottnt{N_{{\mathrm{2}}}}$, respectively.  We
identify two $\alpha$-equivalent terms as usual.  The \emph{size} of a
term is the number of its AST.
%
%
%
%

A typing context $\Gamma$ is a set of bindings of the form
$\mathit{x}\mathord{:}\ottnt{T}$.  It can be seen a partial function from
variables to types.
The typing judgment is of the form $\Gamma  \vdash  \ottnt{M} \,  \mathrel{:}  \, \ottnt{T}$ and asserts that
$\ottnt{M}$ has type $\ottnt{T}$ under the context $\Gamma$; the Curry--Howard
isomorphism allows it to be seen as a judgment asserting $\ottnt{M}$ is a
proof of $\ottnt{T}$ under the assumptions in $\Gamma$.
\fig{typing} shows the typing rules.
Holes may have any type (\T{Hole}); the other rules are standard.
%
%


\section{Sequence-to-sequence neural network}
\label{sec:nn}

We use the \emph{sequence-to-sequence ({\seqseq}) network} as a neural network
to translate a type to its inhabitant.
This section reviews {\seqseq} briefly; interested readers are
referred to Sutskever et al.~\cite{DBLP:journals/corr/SutskeverVL14}
for details.  We also assume basic familiarity about how a neural
network conducts an inference and how a neural network is trained.

In general, application of a (deep) neural network to a supervised
learning problem consists of two phases: training and inference.
The goal of the training phase is to generate a model that
approximates the probability distribution of a given dataset called
training dataset.
In a supervised learning problem, a training dataset consists of pairs
of an input to and an expected output from the trained DNN model.
For example, training for an image recognition task approximates
likelihood of classes of images by taking images as inputs and their
classes as expected outputs.

Training {\seqseq} model approximates conditional probability
distribution $p(y_1,...,y_m \mid x_1,...,x_n)$ where $x_1, ..., x_n$
is an input and $y_1, ..., y_m$ is an output sequence.
After training, the trained model can be used to predict sequences
from inputs in the inference phase.

\begin{figure}[t] \centering
 \includegraphics[width=\textwidth,clip]{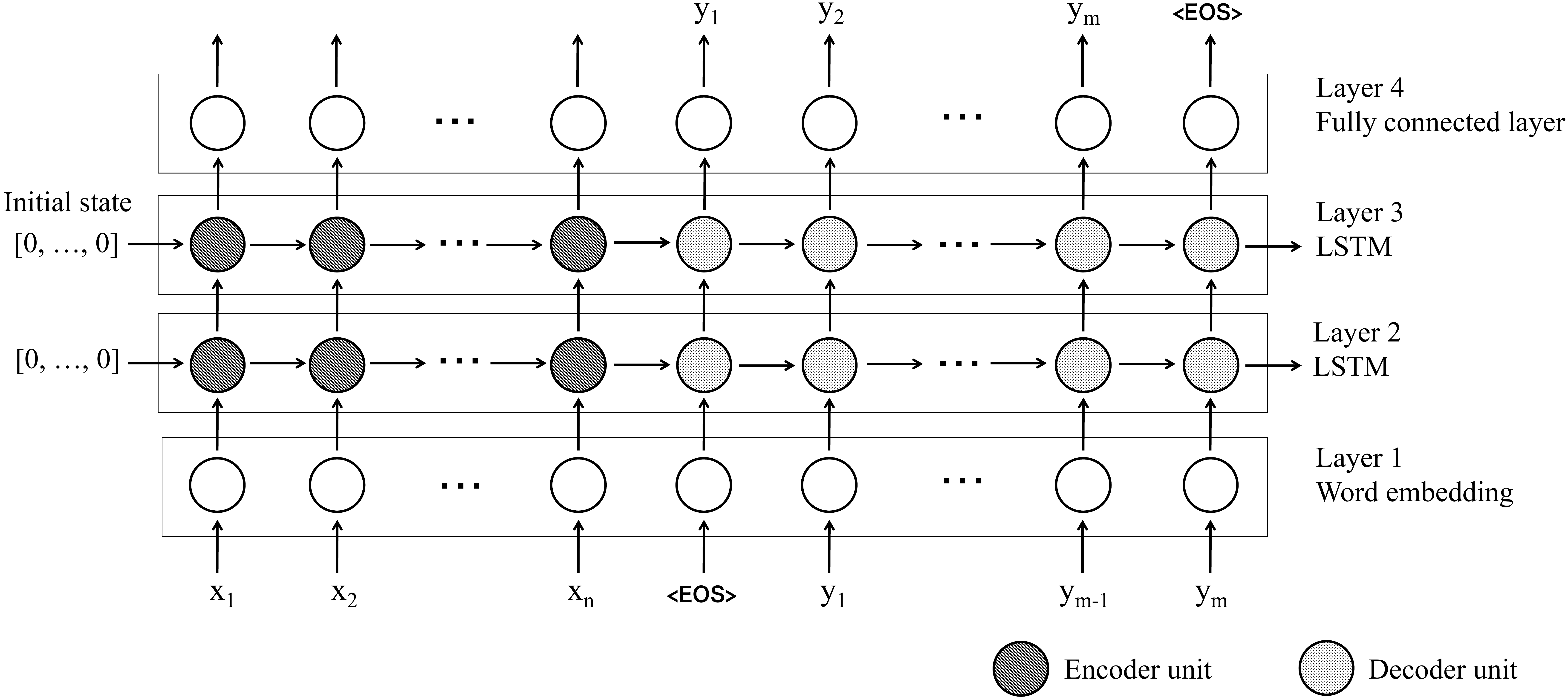}
 \caption{{\seqseq} that takes input sequence $(x_1, \dots, x_n)$ and produces
 output sequence $(y_1, \dots, y_n)$.}
 \label{fig:seq2seq}
\end{figure}
\begin{figure}[t] \centering
 \includegraphics[width=.8\textwidth,clip]{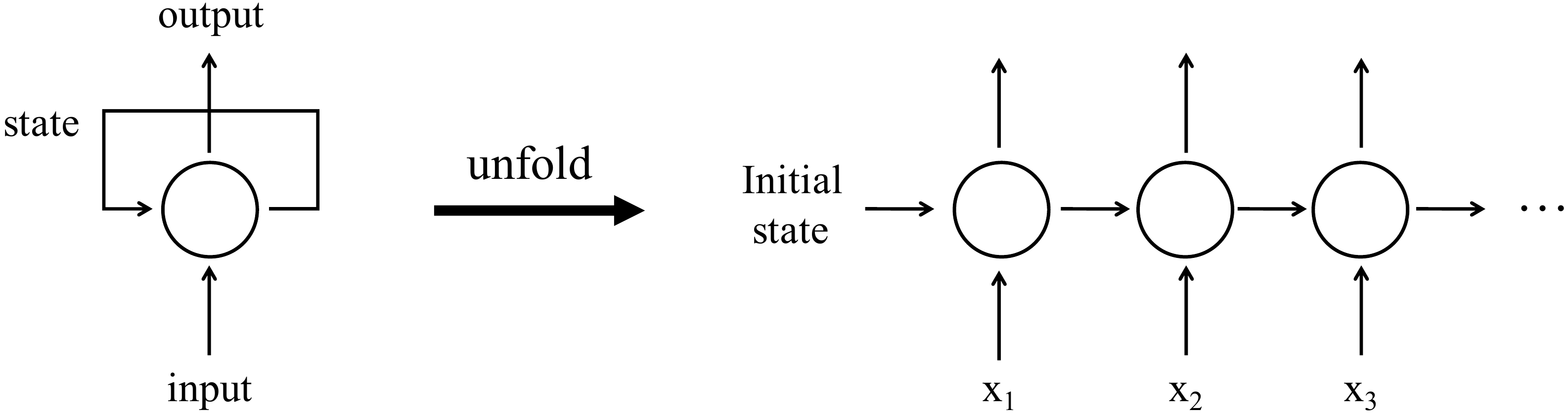}
 \caption{Unfolding a LSTM unit for input sequence $(x_1, x_2, x_3, \dots)$.}
 \label{fig:lstm}
\end{figure}
An overview of the inference with a {\seqseq} model is shown in
\fig{seq2seq}, where $X = (x_1, \dots, x_n)$ is an input and $Y =
(y_1, \dots, y_m)$ is an output sequence.
For each $x_i$, {\seqseq} performs the following procedure.
\begin{enumerate}
 \item $x_i$ is converted to a one-hot vector, which is a $1 \times n$
   matrix ($n$ is the number of vocabularies used in a dataset) where
   all cells are $0$ except that the cell for $x_i$ is $1$.

 \item The one-hot vector is converted to a matrix by the word
       embedding~\cite{DBLP:journals/jmlr/BengioDVJ03,DBLP:journals/corr/abs-1301-3781}
       (Layer 1), which compresses sparse, high-dimensional vector
       representations of words to dense, low-dimensional matrices.

 \item The output of word embedding is processed by 2 Long Short-Term Memory
       (LSTM) units~\cite{DBLP:journals/neco/HochreiterS97} (Layers 2--3).
       LSTMs form a directed circular graph and will be unfolded by following
       the length of an input sequence, as in \fig{lstm}.
       They take an input and the previous state, which is a matrix that has the
       information of the past inputs, and apply matrix operations to produce
       the output and the state for the next input.
       LSTMs can conduct a stateful inference; future outputs can depend
       on past inputs.  This property is important for learning with
       time-series data such as sentences.
       In our system, the initial state is the zero vector.

 \item Finally, the output from the second LSTM is converted to a vector with
       $n$ elements at the fully connected layer (Layer 4), and the vector is
       translated to a token that is most likely.
\end{enumerate}

In Figure~\ref{fig:seq2seq}, the snapshot of a model at an instant is aligned
vertically.  These snapshots are aligned horizontally from left to right along
the flow of time.  An input to a {\seqseq} model is a sequence of data $x_1,
\dots, x_n$, each of which is encoded as a one-hot vector.
The input is terminated with a special symbol \texttt{<EOS>}, which means the end
of the sequence.  The response from the model for the symbol \texttt{<EOS>} is set
to the first element $y_1$ of the output sequence.  An output element $y_i$ is fed
to the model to obtain the next output element $y_{i+1}$ until the model produces
\texttt{<EOS>}.

%

The LSTM layers work as \emph{encoders} while the model is fed with
an input sequence $x_1, \dots, x_n$.  They work as \emph{decoders}
after the model receives \texttt{<EOS>} and while the model produces
an output sequence $y_1,\dots,y_n$.

Since inputs to and outputs from a {\seqseq} model are sequences, in
order to apply {\seqseq} to the proof-synthesis problem, we need a
sequence representation of a type.
%
As the sequence representation of a type $\ottnt{T}$, we use the token sequence
provided by a Haskell interpreter GHCi; this representation is written
$\gamma \, \ottsym{(}  \ottnt{T}  \ottsym{)}$.
For example, $\gamma \, \ottsym{(}  \alpha \,  \mathord{\rightarrow}  \, \ottsym{(}  \alpha \,  \mathord{\rightarrow}  \, \beta  \ottsym{)} \,  \mathord{\rightarrow}  \, \beta  \ottsym{)}$ is ( ``$\alpha$'', ``$ \rightarrow $'',
  ``('', ``$\alpha$'', ``$ \rightarrow $'', ``$\beta$'', ``)'', ``$ \rightarrow $'',
  ``$\beta$'').
This choice of the token-sequence representation is for convenience of
implementation; since we use GHCi as a type checker, using
token sequences in GHCi is convenient.
We train {\seqseq} so that, when it is fed with the GHCi
token-sequence representation of a type, it outputs the token-sequence
representation of a GHCi term.
%
%
%
%
%
%
We also write $\gamma \, \ottsym{(}  \ottnt{M}  \ottsym{)}$ for the GHCi token-sequence representation
of term $\ottnt{M}$.

Transforming outputs from {\seqseq} to terms is a tricky part because
an output of a {\seqseq} model is not always parsable as a term, as we
see in Section~\ref{sec:exp}.
Our synthesis procedure in \sect{synthesis} corrects parsing errors
and finds the nearest parsable token sequence by Myers'
algorithm~\cite{DBLP:journals/ipl/Myers95}.
%
%

We hereafter use a metavariable $\mathcal{M}$ for a trained {\seqseq}
model; we write $\mathcal{M}  \ottsym{(}  \mathcal{S}  \ottsym{)}$ for the sequence that $\mathcal{M}$
infers from the input sequence $\mathcal{S}$.

\section{Program Synthesis Procedure}
\label{sec:synthesis}


\begin{procedure}[t]
  \caption{Synthesis}
  \label{program_synthesis}
  \begin{algorithmic}[1]
    \Procedure{Synthesis}{$T$}
    \State $\mathcal{S} \gets \mathcal{M}  \ottsym{(}  \gamma \, \ottsym{(}  \ottnt{T}  \ottsym{)}  \ottsym{)}$ \label{synth:line2}
    \Comment{Feed ${\seqseq}$ with the given type}
    \State $\ottnt{M} \gets \Call{NearestTerm}{\mathcal{S}}$ \label{synth:line3}
    \Comment{Parse $\mathcal{S}$ and obtain a \emph{guide term}}
    \State $q \gets$ The empty heap of closed terms of $\ottnt{T}$ (ordered by $\COST{\ottnt{M}}{-}$) \label{synth:searchStart}
    \State $\Call{Push}{q,  \left[ \, \right] }$
    \Comment{Proof search starts with the hole term}
    \Loop \Comment{Search guided by $M$}
      \Repeat
        \State $\ottnt{N} \gets \Call{ExtractMin}{q}$
      \Until {find term $\ottnt{N}$ that has not been investigated yet}
      \If {$\ottnt{N}$ contains no holes}
        \State \Return $\ottnt{N}$
      \Else
        \ForEach {$\ottnt{N'} \in \NextTerms(\ottnt{N}, \ottnt{T})$}
          \State $\Call{Push}{q, \ottnt{N'}}$
        \EndFor
      \EndIf
    \EndLoop \label{bfs} \label{synth:searchEnd}
    \EndProcedure
  \end{algorithmic}
\end{procedure}

Procedure~\ref{program_synthesis} is the pseudocode of the procedure
$\Call{Synthesis}{T}$ that takes a type $\ottnt{T}$ and generates a closed,
hole-free term $\ottnt{M}$ of $\ottnt{T}$.
The procedure $\Call{Synthesis}{T}$ uses a trained {\seqseq} model
$\mathcal{M}$; it is in advance trained to generate an inhabitant of a
type.

Line~\ref{synth:line2} feeds $\mathcal{M}$ with $\gamma \, \ottsym{(}  \ottnt{T}  \ottsym{)}$ and obtains
a token sequence $\mathcal{S}$ that is expected to be close to an
inhabitant of $\ottnt{T}$.  This generated $\mathcal{S}$ may be incorrect in
the following two senses: (1) it may not be parsable (i.e., there may
not be $\ottnt{M}$ such that $\gamma \, \ottsym{(}  \ottnt{M}  \ottsym{)} = \mathcal{S}$) and (2) even if such
$\ottnt{M}$ exists, $\ottnt{M}$ may not be an inhabitant of $\ottnt{T}$.
$\Call{Synthesis}{\ottnt{T}}$ fills these two gaps by postprocessing
the output $\mathcal{S}$ from {\seqseq} with the following two computations:
\begin{description}
\item[Guide synthesis] Line~\ref{synth:line3} calls procedure
  $\textsc{NearestTerm}$ that computes $\ottnt{M}$ such that the edit
  distance between $\gamma \, \ottsym{(}  \ottnt{M}  \ottsym{)}$ and $\mathcal{S}$ is smallest.
  $\textsc{NearestTerm}$ uses Myers'
  algorithm~\cite{DBLP:journals/ipl/Myers95}.  The output term $\ottnt{M}$ from
  $\textsc{NearestTerm}$ is called a \emph{guide term}.
\item[Guided proof search]
  Lines~\ref{synth:searchStart}--\ref{synth:searchEnd} enumerate
  candidate terms and test whether each candidate is a proof term of
  $\ottnt{T}$.  In order to give higher priority to a candidate term
  that is ``closer'' to guide term $\ottnt{M}$, the procedure designates a priority
  queue $q$.  This queue orders the enqueued terms by the value of a
  cost function $\COST{\ottnt{M}}{-}$.  The cost function is a parameter
  of the procedure $\Call{Synthesis}{\ottnt{T}}$; it is defined so that
  the value of $\COST{\ottnt{M}}{\ottnt{M'}}$ is smaller if the tree edit
  distance~\cite{DBLP:journals/siamcomp/ZhangS89} between $\ottnt{M}$ and $\ottnt{M'}$
  is smaller.  We present the definition of the cost functions that we
  use later.  The enumeration of the candidate terms is continued
  until $\Call{Synthesis}{\ottnt{T}}$ encounters a correct proof of
  $\ottnt{T}$.  Although it is not guaranteed that this procedure
  converges,\footnote{If $\ottnt{T}$ does not have an inhabitant, then
    $\Call{Synthesis}{\ottnt{T}}$ indeed diverges.} experiments presented in
  \sect{exp-synthesis} indicate that
  $\Call{Synthesis}{\ottnt{T}}$ discovers a proof fast in many cases
  compared to a brute-force proof search.
  
\end{description}

\begin{figure}[t]
 \[\begin{array}{lll}
 \mathcal{C} &::=& \mathit{x} \mid \lambda  \mathit{x}  \ottsym{.}  \left[ \, \right] \mid \left[ \, \right] \, \left[ \, \right] \mid
         \ottsym{(}  \left[ \, \right]  \ottsym{,}  \left[ \, \right]  \ottsym{)} \mid \mathsf{case} \, \left[ \, \right] \, \mathsf{of} \, \ottsym{(}  \mathit{x}  \ottsym{,}  \mathit{y}  \ottsym{)}  \rightarrow  \left[ \, \right] \mid \\ &&
         \mathsf{Left} \, \left[ \, \right] \mid \mathsf{Right} \, \left[ \, \right] \mid
         \mathsf{case} \, \left[ \, \right] \, \mathsf{of} \, \ottsym{\{} \, \mathsf{Left} \, \mathit{x}  \rightarrow  \left[ \, \right]  \ottsym{;} \, \mathsf{Right} \, \mathit{y}  \rightarrow  \left[ \, \right]  \ottsym{\}} \\
 \end{array}\]
 \caption{Shallow contexts.}
 \label{fig:shallow}
\end{figure}

\sloppy{
The remaining ingredient of the guided proof-search phase
(Lines~\ref{synth:searchStart}--\ref{synth:searchEnd}) is the
subprocedure $\Call{GenCandidates}{\ottnt{N},\ottnt{T}}$ that generates
candidate terms.  $\Call{GenCandidates}{\ottnt{N},\ottnt{T}}$ takes two
parameters: term $\ottnt{N}$ which may contain several holes and type
$\ottnt{T}$ of candidate terms to be synthesized.
$\Call{GenCandidates}{\ottnt{N},\ottnt{T}}$ generates the set of the terms
that are obtained by filling a hole in $\ottnt{N}$ with a \emph{shallow
  context} $\mathcal{C}$, a depth-1 term with holes, which is defined
  in Figure~\ref{fig:shallow}.  Concretely, $\Call{GenCandidates}{\ottnt{N},\ottnt{T}}$
constructs a set of candidate terms by the following steps: (1)
constructing the set $S$ such that $\ottnt{N'} \in S$ if and only if $\ottnt{N'}$ is
obtained by filling a hole in $N$ with an arbitrary shallow context
$\mathcal{C}$ in which only the variables bound at the hole can occur
freely;\footnote{Since we identify $\alpha$-equivalent terms, the
  number of the shallow contexts that can be filled in is finite.} and
(2) filtering out from $S$ the terms that contain a $\beta\eta$-redex (to prune
the proof-search space) or do not have type $\ottnt{T}$.
}






\section{Preliminary Experiments}
\label{sec:exp}



In order to confirm the baseline of the proof synthesis with {\seqseq} and the
feasibility of our proof-synthesis method, we train {\seqseq} models,
implement $\Call{Synthesis}{T}$, and measure the performance of the
implemented procedure.

\subsection{Environment}
\label{sec:environment}

%

We implemented {\SYNTHESIS} in Python 3 (version 3.6.1) except for
\textsc{NearestTerm}, which is implemented with OCaml 4.04.1, and the type
checker, for which we use Haskell interpreter GHCi 8.0.1; we write
$\textit{TypeInf} \, \ottsym{(}  \ottnt{M}  \ottsym{)}$ for the type of $\ottnt{M}$ inferred by the GHCi.
Training of {\seqseq} and inference using the trained {\seqseq} models are
implemented with a deep learning framework Chainer (version
1.24.0)~\cite{tokui2015chainer};\footnote{We used the code available at
\url{https://github.com/kenkov/seq2seq} with an adaptation.} as the word2vec
module, we used one provided by Python library gensim (version 0.13.4).
We conduct all experiments on an instance provided by Amazon EC2
g2.2xlarge, equipped with 8 virtual CPUs (Intel Xeon E5-2670; 2.60
GHz, 4 cores) and 15 GiB memory.
Although the instance is equipped with a GPU, it is not used in the
training phase nor the inference phase.


As shown in \fig{seq2seq}, our {\seqseq} network consists of
4 layers: a layer for word embedding, 2 LSTMs, and a fully connected
layer.
\begin{table}[t]
 \begin{center}
  \begin{tabular}{|@{\ }c@{\ }|@{\ }c@{\ }|} \hline
   Layer type            & The number of parameters \\ \hline \hline
   Word embedding        & $W \times 50$ \\ \hline
   LSTM                  & 20 K \\ \hline
   LSTM                  & 20 K \\ \hline
   Fully connected layer & $W \times 50$ \\ \hline
  \end{tabular}
 \end{center}
 \caption{Learnable parameters in the network: $W$ is the number of
   the vocabularies.}
 \label{tbl:param}
\end{table}
Their learnable parameters are shown in \tbl{param}, where $W$ is the
number of vocabularies used in the dataset.  The value of $W$ depends
on the token-sequence representation of training data; in the current
dataset, it is $40$.
The parameters in the word embedding are initialized by
\texttt{word2vec}~\cite{DBLP:journals/corr/abs-1301-3781}; we used
CBOW with negative sampling; the window size is set to 5.
The weights of the LSTMs and the last fully connected layer are
initialized by i.i.d.\ Gaussian sampling with mean 0 and deviation
$\sqrt{50}$ (the number 50 is the output size of the previous layer of
each); the biases are initialized to 0.
We trained the model with stochastic gradient descent.
The loss function is the sum of the softmax cross entropy
accumulated over the token sequence.
As an optimizer, we use Adam~\cite{DBLP:journals/corr/KingmaB14} with the
following parameters: $\alpha = 0.001$, $\beta_1 = 0.9$, $\beta_2 = 0.999$, and
$\epsilon = 10^{-8}$.

\subsection{Generating dataset}

\begin{procedure}[t]
  \caption{Generation of training dataset} \label{proc:train}
  \begin{algorithmic}[1]
    \Procedure{TrainingDataset}{$n$}
       \Comment{Make $n$ pairs of a type and its term}
    \State $dataset \gets \{\}$
    \While{$size(dataset) < n$}
      \label{train:random_size}
      \State $s \gets$ choose from $\{2, ..., 9\}$ at uniformly random
      \label{train:random_term}
      \State $\ottnt{M}$ $\gets$ generate a closed, hole-free, well-typed term of size $s$
             at random
      \label{train:add_start}
      \State $\ottnt{T}$ $\gets$ $\textit{TypeInf} \, \ottsym{(}  \ottnt{M}  \ottsym{)}$
      \If {$(\ottnt{T}, \ottnt{N}) \in dataset$ for some $\ottnt{N}$}
        \If {$\textit{size} \, \ottsym{(}  \ottnt{M}  \ottsym{)} < \textit{size} \, \ottsym{(}  \ottnt{N}  \ottsym{)}$}
          \State $dataset \gets (dataset \mathbin{\backslash} \{(\ottnt{T},\ottnt{N})\}) \mathbin{\cup} \{(\ottnt{T}, \ottnt{M})\}$
        \EndIf
      \Else
        \State $dataset \gets dataset \mathbin{\cup} \{(\ottnt{T}, \ottnt{M})\}$
      \label{train:add_end}
      \EndIf
    \EndWhile
    \State \Return $dataset$
    \EndProcedure
  \end{algorithmic}
\end{procedure}

In order to train the network, we need a set of type--term pairs as
training data.  Since we are not aware of publicly available such
data, we used data generated by
Procedure~\ref{proc:train} for this purpose.  This procedure first
uniformly chooses an integer $s$ from 1 to 9, uniformly samples a term
$\ottnt{M}$ from the set of the size-$s$ terms, and adds it to the
dataset.\footnote{We also conducted an experiment with a dataset that
  only consists of $\beta\eta$-normal terms; see \sect{exp-model}.}
If $dataset$ already contains a term $\ottnt{N}$ of the type
$\textit{TypeInf} \, \ottsym{(}  \ottnt{M}  \ottsym{)}$ of $\ottnt{M}$, the smaller one is assigned to $\ottnt{T}$;
otherwise, $\ottnt{M}$ is added to $dataset$.  Models are trained on a
training set that consists of 1000 pairs of a type $\ottnt{T}$ and a
closed hole-free term $\ottnt{M}$ of $\ottnt{T}$.

We do not claim that a dataset generated by $\Call{TrainingDataset}{n}$
is the best for training.  Ideally, a training dataset should reflect
the true distribution of the type--term pairs.  The current dataset
does not necessarily reflect this distribution in that (1) it ignores
a proof with size more than $9$ and (2) it gives higher preference to
smaller proofs.\footnote{We observed that the number of well-typed
  terms grows exponentially with respect to the size of a term;
  therefore, if we uniformly sample training data from the set of
  well-typed terms, a term with smaller size is rarely included in the
  dataset.  By first fixing the size $s$ and then uniformly sampling
  the term of size $s$, we give relatively higher preference to
  smaller-size terms.}  By using the repository of hand-written
proofs as the training dataset, we could approximate this
distribution, which is an important future direction of our work.

\subsection{Training}
\label{sec:exp-model}

We train the network using two datasets: $D_\textit{tm}$ generated by
Procedure~\ref{proc:train} and $D_{\beta\eta}$ generated in the same
way but contains only $\beta\eta$ normal forms.
%
%
We trained the network not only with $D_{tm}$ but with $D_{\beta\eta}$
because a proof term with a $\beta\eta$-redex can be seen as a detour
from the viewpoint of the proof
theory~\cite{Sorensen:2006:LCI:1197021}; therefore, we expect that the
model trained with $D_{\beta\eta}$ generates a guide term that is more
succinct than one trained with $D_\textit{tm}$.
%
%
We used the following batch sizes in the training in each training
dataset: 16, 32, and 64.
Each model is trained for 3000 epochs.
\begin{figure}[t]
 \includegraphics[width=.5\textwidth,clip]{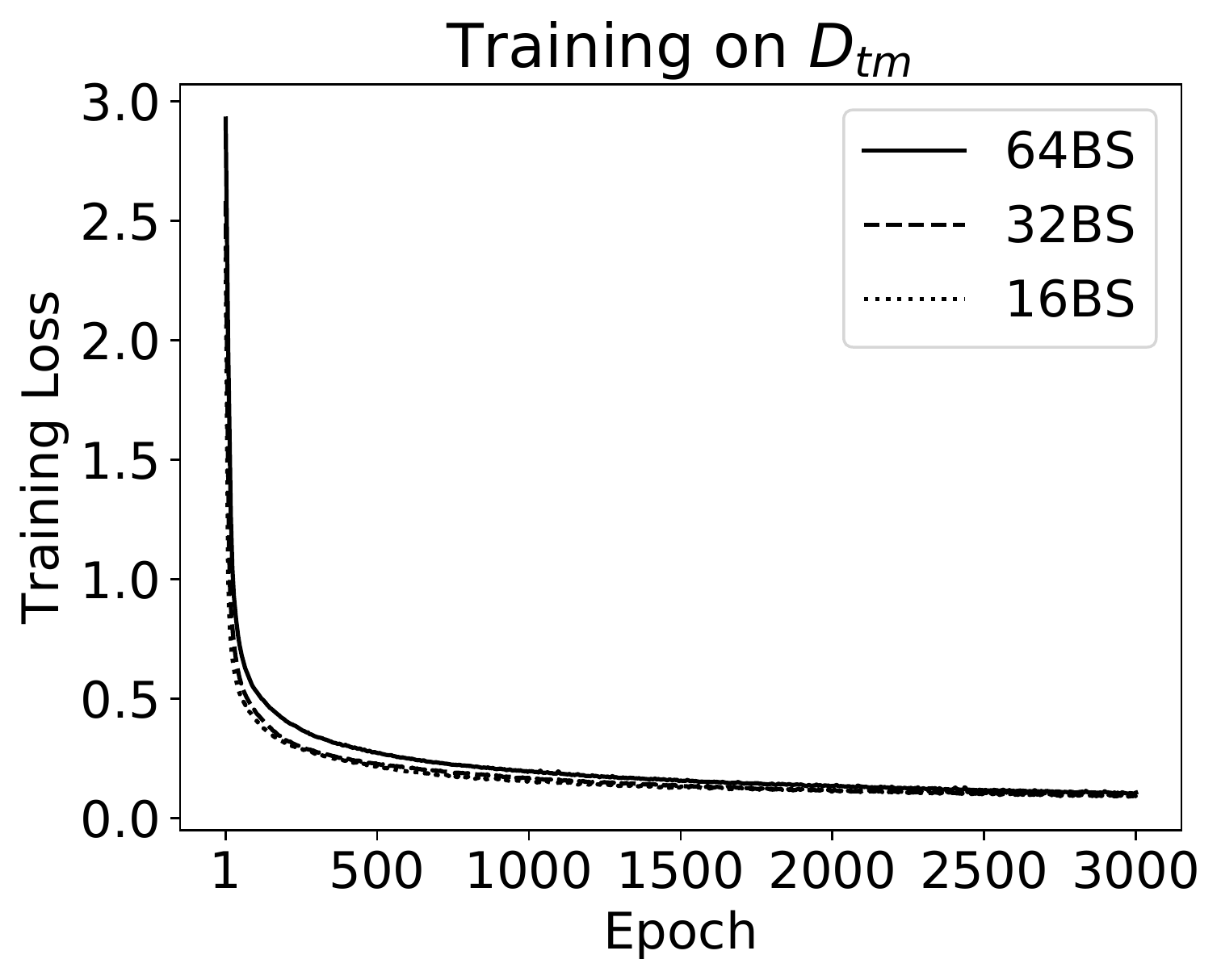}
 \includegraphics[width=.5\textwidth,clip]{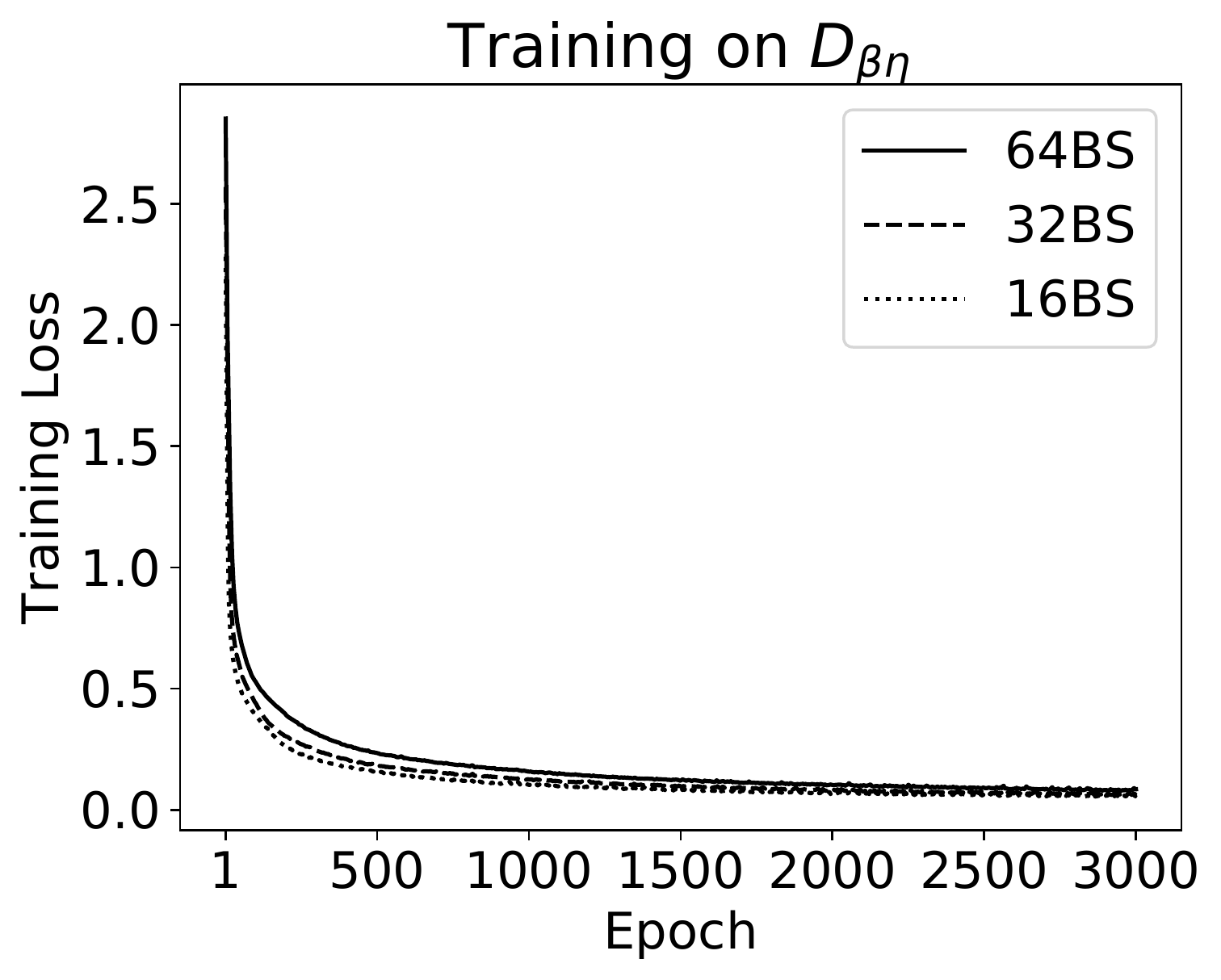}
 \caption{Smoothed plots of the training loss over epochs: the left
   graph shows the plots for the models trained with $D_\textit{tm}$;
   the right graph shows the plots for the models trained with
   $D_{\beta\eta}$; each graph contains the plots for different batch
   sizes (BS).}
 \label{fig:train-loss}
\end{figure}

\fig{train-loss} shows the smoothed plots of the training loss over epochs.
Since a loss represents the difference between expected outputs and
actual outputs from the trained model, these graphs mean that the training of
each model almost converges after 3000 epochs.
%

\begin{table}[t]
 \begin{center}
 \begin{tabular}{|c|c||c|} \hline
  \multicolumn{2}{|c||}{Model} &
  \multirow{2}{*}{Inferred term from $\alpha_{{\mathrm{1}}} \,  \mathord{\times}  \, \alpha_{{\mathrm{2}}} \,  \mathord{\rightarrow}  \, \alpha_{{\mathrm{2}}} \,  \mathord{\times}  \, \alpha_{{\mathrm{1}}}$} \\ \cline{1-2}
  Dataset & Batch size & \\ \hline \hline
  $D_\textit{tm}$ & 16 &
   $\ottsym{(}  \lambda  \mathit{x_{{\mathrm{0}}}}  \ottsym{.}  \ottsym{(}  \mathsf{case} \, \mathit{x_{{\mathrm{0}}}} \, \mathsf{of} \, \ottsym{(}  \mathit{x_{{\mathrm{1}}}}  \ottsym{,}  \mathit{x_{{\mathrm{2}}}}  \ottsym{)}  \rightarrow  \ottsym{(}  \mathit{x_{{\mathrm{1}}}}  \ottsym{,}  \mathit{x_{{\mathrm{1}}}}  \ottsym{)}  \ottsym{)}  \ottsym{)}$
   \\ \hline
  $D_\textit{tm}$ & 32 &
   $\ottsym{(}  \lambda  \mathit{x_{{\mathrm{0}}}}  \ottsym{.}  \ottsym{(}  \mathsf{case} \, \mathit{x_{{\mathrm{0}}}} \, \mathsf{of} \, \ottsym{(}  \mathit{x_{{\mathrm{1}}}}  \ottsym{,}  \mathit{x_{{\mathrm{2}}}}  \ottsym{)}  \rightarrow  \ottsym{(}  \mathit{x_{{\mathrm{1}}}}  \ottsym{,}  \mathit{x_{{\mathrm{1}}}}  \ottsym{)}  \ottsym{)}  \ottsym{)}$
   \\ \hline
  $D_\textit{tm}$ & 64 &
   $\ottsym{(}  \lambda  \mathit{x_{{\mathrm{0}}}}  \ottsym{.}  \ottsym{(}  \ottsym{(}  \mathsf{case} \, \mathit{x_{{\mathrm{0}}}} \, \mathsf{of} \, \ottsym{(}  \mathit{x_{{\mathrm{1}}}}  \ottsym{,}  \mathit{x_{{\mathrm{2}}}}  \ottsym{)}  \rightarrow  \mathit{x_{{\mathrm{1}}}}  \ottsym{)}  \ottsym{,}  \ottsym{(}  \mathsf{Left} \, \mathit{x_{{\mathrm{0}}}}  \ottsym{)}  \ottsym{)}  \ottsym{)}$
   \\ \hline
  $D_{\beta\eta}$ & 16 &
   $\ottsym{(}  \lambda  \mathit{x_{{\mathrm{0}}}}  \ottsym{.}  \ottsym{(}  \ottsym{(}  \mathsf{case} \, \mathit{x_{{\mathrm{0}}}} \, \mathsf{of} \, \ottsym{(}  \mathit{x_{{\mathrm{1}}}}  \ottsym{,}  \mathit{x_{{\mathrm{2}}}}  \ottsym{)}  \rightarrow  \mathit{x_{{\mathrm{1}}}}  \ottsym{)}  \ottsym{,}  \mathit{x_{{\mathrm{0}}}}  \ottsym{)}  \ottsym{)}$
   \\ \hline
  $D_{\beta\eta}$ & 32 &
   $\ottsym{(}  \lambda  \mathit{x_{{\mathrm{0}}}}  \ottsym{.}  \ottsym{(}  \mathsf{case} \, \mathit{x_{{\mathrm{0}}}} \, \mathsf{of} \, \ottsym{(}  \mathit{x_{{\mathrm{1}}}}  \ottsym{,}  \mathit{x_{{\mathrm{2}}}}  \ottsym{)}  \rightarrow  \ottsym{(}  \mathit{x_{{\mathrm{1}}}}  \ottsym{,}  \mathit{x_{{\mathrm{1}}}}  \ottsym{)}  \ottsym{)}  \ottsym{)}$
   \\ \hline
  $D_{\beta\eta}$ & 64 &
   $\ottsym{(}  \lambda  \mathit{x_{{\mathrm{0}}}}  \ottsym{.}  \ottsym{(}  \mathsf{case} \, \mathit{x_{{\mathrm{0}}}} \, \mathsf{of} \, \ottsym{(}  \mathit{x_{{\mathrm{1}}}}  \ottsym{,}  \mathit{x_{{\mathrm{2}}}}  \ottsym{)}  \rightarrow  \ottsym{(}  \mathit{x_{{\mathrm{1}}}}  \ottsym{,}  \mathit{x_{{\mathrm{0}}}}  \ottsym{)}  \ottsym{)}  \ottsym{)}$
   \\ \hline
 \end{tabular} \vspace{0.5cm}

 \end{center}

 \caption{Examples of terms inferred by trained {\seqseq} models.}
 \label{tbl:inferred-term}
\end{table}

\sloppy{
\tbl{inferred-term} shows examples of terms inferred by trained models from type
$\alpha_{{\mathrm{1}}} \,  \mathord{\times}  \, \alpha_{{\mathrm{2}}} \,  \mathord{\rightarrow}  \, \alpha_{{\mathrm{2}}} \,  \mathord{\times}  \, \alpha_{{\mathrm{1}}}$, which denotes swapping components of pairs.
All terms shown in \tbl{inferred-term} capture that they should take a pair
($\mathit{x_{{\mathrm{0}}}}$), decompose it by a case expression, and compose a new pair using the
decomposition result.
Unfortunately, they are \emph{not} correct proof terms of the
designated type: terms in the first, second, fifth, and sixth rows
refer to incorrect variables; and ones in the third and fourth rows
decompose the argument pair only for making the first element of the
new pair.
On the other hand, they somewhat \emph{resemble} to a correct proof
term (e.g., $\lambda  \mathit{x_{{\mathrm{0}}}}  \ottsym{.}  \mathsf{case} \, \mathit{x_{{\mathrm{0}}}} \, \mathsf{of} \, \ottsym{(}  \mathit{x_{{\mathrm{1}}}}  \ottsym{,}  \mathit{x_{{\mathrm{2}}}}  \ottsym{)}  \rightarrow  \ottsym{(}  \mathit{x_{{\mathrm{2}}}}  \ottsym{,}  \mathit{x_{{\mathrm{1}}}}  \ottsym{)}$).  Our synthesis
procedure uses a generated (incorrect) proof term to efficiently
search for a correct one.}

\subsection{Evaluation of the trained models}
\label{sec:evaluateModels}
We quantitatively evaluate our models on the following aspects:
\begin{description}
\item[Parsability] How many inferred strings are parsable as proof
  terms?
\item[Typability] How many inferred terms, after the postprocessing by
  $\textsc{NearesrtTerm}$, are indeed correct proofs?
%
%
%
\item[Closeness] How close is an inferred proof postprocessed by
  $\textsc{NearesrtTerm}$ to a correct proof in average?
%
%
\end{description}
We measure the closeness by tree edit
distance~\cite{DBLP:journals/siamcomp/ZhangS89};
%
we measure the edit distance between an inferred term and a correct
proof term that is the closest to the inferred term and whose size is
not more than $9$.

We generated terms using the trained models with a test dataset that consists of
1000 types sampled by the similar way to Procedure~\ref{proc:train} but does not
contain any type in $D_\textit{tm}$ nor $D_{\beta\eta}$.
The evaluation results of the quantities above are shown in \tbl{eval-model}.
We discuss the result below.
\begin{table}[t]
 \begin{center}
 \begin{tabular}{|@{\,}c@{\,}|l@{\,}|c|c|c|c|c|c|} \hline
  \multirow{2}{*}{Model} & \ Dataset & $D_\textit{tm}$ & $D_\textit{tm}$ & $D_\textit{tm}$ & $D_{\beta\eta}$ & $D_{\beta\eta}$ & $D_{\beta\eta}$ \\ \cline{2-8}
  & \ Batch size & 16 & 32 & 64 & 16 & 32 & 64 \\ \hline \hline
  \multirow{7}{*}{Evaluation} & \ \# of parsable & 983 & 987 & 987 & 991 & 988 & 990 \\ \cline{2-8}
  & \ \# of typable & 430 & 510 & 463 & 515 & 475 & 451 \\ \cline{2-8}
  & \ Rate of misuse of vars (\%)& 39.82 & 30.61 & 42.27 & 28.45 & 33.90 & 30.78 \\ \cline{2-8}
  & \ Closeness per AST node & 0.1982 & 0.1805 & 0.1831 & 0.1878 & 0.1822 & 0.2001 \\ \hline
 \end{tabular}
 \end{center}

 \caption{Evaluation of the trained models.}
 \label{tbl:eval-model}
\end{table}
\begin{itemize}
\item Every model generates a parsable response in more than 980
  propositions out of 1000.  This rate turns out to be high enough for
  the call to $\textsc{NearestTerm}$ in the synthesis procedure to
  work in reasonable time.
\item As for the number of typable responses, the number is between
  430 to 515 out of 1000 depending on the training data and the batch
  size.  We observed that the error is often caused due to the misuse
  of variables.  For example, as shown in \tbl{inferred-term},
  $\ottsym{(}  \lambda  \mathit{x_{{\mathrm{0}}}}  \ottsym{.}  \ottsym{(}  \mathsf{case} \, \mathit{x_{{\mathrm{0}}}} \, \mathsf{of} \, \ottsym{(}  \mathit{x_{{\mathrm{1}}}}  \ottsym{,}  \mathit{x_{{\mathrm{2}}}}  \ottsym{)}  \rightarrow  \ottsym{(}  \mathit{x_{{\mathrm{1}}}}  \ottsym{,}  \mathit{x_{{\mathrm{1}}}}  \ottsym{)}  \ottsym{)}  \ottsym{)}$ is inferred
  as a proof term for the proposition $\alpha_{{\mathrm{1}}} \,  \mathord{\times}  \, \alpha_{{\mathrm{2}}} \,  \mathord{\rightarrow}  \, \alpha_{{\mathrm{2}}} \,  \mathord{\times}  \, \alpha_{{\mathrm{1}}}$.
  Although this term is incorrect, this term is made correct by
  replacing the first reference to $\mathit{x_{{\mathrm{1}}}}$ with that to $\mathit{x_{{\mathrm{2}}}}$.
  The row ``Rate of misuse of vars'' in \tbl{eval-model} is the rate
  of such errors among the whole erroneous terms.  Given such errors
  are frequent, we guess that the combination of our method and
  premise selection heuristics is promising in improving the accuracy.
\item Closeness is measured by the average of the per-node minimum tree edit
  distance between a generated term (postprocessed by
  $\textsc{NearestTerm}$) and a correct proof term whose sizes are not
  more than $9$.  The precise definition is
    \[
    \frac{1}{1000}\sum_{i = 1}^{1000}\frac{\min(\{ \textit{EditDist} \, \ottsym{(}  \ottnt{N_{\ottmv{i}}}  \ottsym{,}  \ottnt{M}  \ottsym{)} \mid
      \ottnt{M} \in D_{\ottnt{T_{\ottmv{i}}}} \})}{\textit{size} \, \ottsym{(}  \ottnt{N_{\ottmv{i}}}  \ottsym{)}}
    \]
    where $\ottnt{T_{\ottmv{i}}}$ is a type for the $i$-th test case; $D_{\ottnt{T_{\ottmv{i}}}}$ is a set
    of closed hole-free terms of type $\ottnt{T_{\ottmv{i}}}$ whose sizes are not more than $9$;
    and $\ottnt{N_{\ottmv{i}}} \, \ottsym{=} \, \textsc{NearestTerm} \, \ottsym{(}  \mathcal{M}  \ottsym{(}  \gamma \, \ottsym{(}  \ottnt{T_{\ottmv{i}}}  \ottsym{)}  \ottsym{)}  \ottsym{)}$.  We can observe that
    we need to edit about 19\% of the whole nodes of a term generated by the
    models in average to obtain a correct proof.  We believe
    that this rate can be made less if we tune the network.
\end{itemize}

\subsection{Evaluation of the synthesis procedures}
\label{sec:exp-synthesis}

We evaluate Procedure~\ref{program_synthesis} with several
instantiations of the cost function.  In the definition of the cost
functions, we use auxiliary function $\textit{imitate} \, \ottsym{(}  \ottnt{N}  \ottsym{,}  \ottnt{M}  \ottsym{)}$, which is
defined as follows:
\[\begin{array}{lll}
 \textit{imitate} \, \ottsym{(}  \left[ \, \right]  \ottsym{,}  \ottnt{M}  \ottsym{)} &=& \ottnt{M} \\
 \textit{imitate} \, \ottsym{(}  \lambda  \mathit{x}  \ottsym{.}  \ottnt{N}  \ottsym{,}  \lambda  \mathit{x}  \ottsym{.}  \ottnt{M}  \ottsym{)} &=& \lambda  \mathit{x}  \ottsym{.}  \textit{imitate} \, \ottsym{(}  \ottnt{N}  \ottsym{,}  \ottnt{M}  \ottsym{)} \\
 \textit{imitate} \, \ottsym{(}  \ottnt{N_{{\mathrm{1}}}} \, \ottnt{N_{{\mathrm{2}}}}  \ottsym{,}  \ottnt{M_{{\mathrm{1}}}} \, \ottnt{M_{{\mathrm{2}}}}  \ottsym{)} &=& \textit{imitate} \, \ottsym{(}  \ottnt{N_{{\mathrm{1}}}}  \ottsym{,}  \ottnt{M_{{\mathrm{1}}}}  \ottsym{)} \, \textit{imitate} \, \ottsym{(}  \ottnt{N_{{\mathrm{2}}}}  \ottsym{,}  \ottnt{M_{{\mathrm{2}}}}  \ottsym{)} \\
 \textit{imitate} \, \ottsym{(}  \ottsym{(}  \ottnt{N_{{\mathrm{1}}}}  \ottsym{,}  \ottnt{N_{{\mathrm{2}}}}  \ottsym{)}  \ottsym{,}  \ottsym{(}  \ottnt{M_{{\mathrm{1}}}}  \ottsym{,}  \ottnt{M_{{\mathrm{2}}}}  \ottsym{)}  \ottsym{)} &=& \ottsym{(}  \textit{imitate} \, \ottsym{(}  \ottnt{N_{{\mathrm{1}}}}  \ottsym{,}  \ottnt{M_{{\mathrm{1}}}}  \ottsym{)}  \ottsym{,}  \textit{imitate} \, \ottsym{(}  \ottnt{N_{{\mathrm{2}}}}  \ottsym{,}  \ottnt{M_{{\mathrm{2}}}}  \ottsym{)}  \ottsym{)} \\
 \multicolumn{3}{l}{
  \textit{imitate} \, \ottsym{(}  \mathsf{case} \, \ottnt{N_{{\mathrm{1}}}} \, \mathsf{of} \, \ottsym{(}  \mathit{x}  \ottsym{,}  \mathit{y}  \ottsym{)}  \rightarrow  \ottnt{N_{{\mathrm{2}}}}  \ottsym{,}  \mathsf{case} \, \ottnt{M_{{\mathrm{1}}}} \, \mathsf{of} \, \ottsym{(}  \mathit{x}  \ottsym{,}  \mathit{y}  \ottsym{)}  \rightarrow  \ottnt{M_{{\mathrm{2}}}}  \ottsym{)}
  } \\
  \multicolumn{3}{l}{ \qquad
   = \mathsf{case} \, \textit{imitate} \, \ottsym{(}  \ottnt{N_{{\mathrm{1}}}}  \ottsym{,}  \ottnt{M_{{\mathrm{1}}}}  \ottsym{)} \, \mathsf{of} \, \ottsym{(}  \mathit{x}  \ottsym{,}  \mathit{y}  \ottsym{)}  \rightarrow  \textit{imitate} \, \ottsym{(}  \ottnt{N_{{\mathrm{2}}}}  \ottsym{,}  \ottnt{M_{{\mathrm{2}}}}  \ottsym{)}
  } \\
 \textit{imitate} \, \ottsym{(}  \mathsf{Left} \, \ottnt{N}  \ottsym{,}  \mathsf{Left} \, \ottnt{M}  \ottsym{)} &=& \mathsf{Left} \, \textit{imitate} \, \ottsym{(}  \ottnt{N}  \ottsym{,}  \ottnt{M}  \ottsym{)} \\
 \textit{imitate} \, \ottsym{(}  \mathsf{Right} \, \ottnt{N}  \ottsym{,}  \mathsf{Right} \, \ottnt{M}  \ottsym{)} &=& \mathsf{Right} \, \textit{imitate} \, \ottsym{(}  \ottnt{N}  \ottsym{,}  \ottnt{M}  \ottsym{)} \\
 \multicolumn{3}{l}{
  \textit{imitate} \, \ottsym{(}  \mathsf{case} \, \ottnt{N_{{\mathrm{1}}}} \, \mathsf{of} \, \ottsym{\{} \, \mathsf{Left} \, \mathit{x}  \rightarrow  \ottnt{N_{{\mathrm{2}}}}  \ottsym{;} \, \mathsf{Right} \, \mathit{y}  \rightarrow  \ottnt{N_{{\mathrm{3}}}}  \ottsym{\}}  \ottsym{,} } \\
  \multicolumn{3}{l}{ \qquad \quad \ \,
  \mathsf{case} \, \ottnt{M_{{\mathrm{1}}}} \, \mathsf{of} \, \ottsym{\{} \, \mathsf{Left} \, \mathit{x}  \rightarrow  \ottnt{M_{{\mathrm{2}}}}  \ottsym{;} \, \mathsf{Right} \, \mathit{y}  \rightarrow  \ottnt{M_{{\mathrm{3}}}}  \ottsym{\}}  \ottsym{)} } \\
  \multicolumn{3}{l}{ \qquad
   = \mathsf{case} \, \textit{imitate} \, \ottsym{(}  \ottnt{N_{{\mathrm{1}}}}  \ottsym{,}  \ottnt{M_{{\mathrm{1}}}}  \ottsym{)} \, \mathsf{of} \, \ottsym{\{} \, \mathsf{Left} \, \mathit{x}  \rightarrow  \textit{imitate} \, \ottsym{(}  \ottnt{N_{{\mathrm{2}}}}  \ottsym{,}  \ottnt{M_{{\mathrm{2}}}}  \ottsym{)}  \ottsym{;} } \\
  \multicolumn{3}{l}{ \qquad \qquad \qquad \qquad \qquad \qquad \quad \ \ %
   \mathsf{Right} \, \mathit{y}  \rightarrow  \textit{imitate} \, \ottsym{(}  \ottnt{N_{{\mathrm{3}}}}  \ottsym{,}  \ottnt{M_{{\mathrm{3}}}}  \ottsym{)}  \ottsym{\}}
 } \\
 \textit{imitate} \, \ottsym{(}  \ottnt{N}  \ottsym{,}  \ottnt{M}  \ottsym{)} &=& \ottnt{N} \quad \text{(Otherwise)}
\end{array}\]
The function $\textit{imitate} \, \ottsym{(}  \ottnt{N}  \ottsym{,}  \ottnt{M}  \ottsym{)}$ matches the term $\ottnt{N}$ that contains
holes against the guide term $\ottnt{M}$; if $\ottnt{N}$ is a hole, then
$\ottnt{M}$ is returned as the result.  If the top constructor of $\ottnt{M}$
is different from $\ottnt{N}$, then it returns $\ottnt{N}$.  This function is
used to express different treatment of a hole in computation of tree
edit distance.

We use the following cost functions computed from a candidate term
$\ottnt{N}$ and a guide term $\ottnt{M}$:
\begin{itemize}
\item $\COSTbf{\ottnt{M}}{\ottnt{N}} := \textit{size} \, \ottsym{(}  \ottnt{N}  \ottsym{)}$ that does not take the
  edit distance between $\ottnt{N}$ and $\ottnt{M}$.  Since this function
  ignores the guide term, Procedure~\ref{program_synthesis}
  instantiated with this cost function searches for a correct proof
  term in a brute-force way.


\item $\COSTed{\ottnt{M}}{\ottnt{N}} := \textit{size} \, \ottsym{(}  \ottnt{N}  \ottsym{)} + \textit{EditDist} \, \ottsym{(}  \ottnt{M}  \ottsym{,}  \ottnt{N}  \ottsym{)}$, where
  $\textit{size} \, \ottsym{(}  \ottnt{N}  \ottsym{)}$ is the size of $\ottnt{N}$ and $\textit{EditDist} \, \ottsym{(}  \ottnt{M}  \ottsym{,}  \ottnt{N}  \ottsym{)}$ is the tree
  edit distance between $\ottnt{M}$ and $\ottnt{N}$.
\item $\COSTim{\ottnt{M}}{\ottnt{N}} := \textit{size} \, \ottsym{(}  \ottnt{N}  \ottsym{)} + \textit{EditDist} \, \ottsym{(}  \ottnt{M}  \ottsym{,}  \textit{imitate} \, \ottsym{(}  \ottnt{N}  \ottsym{,}  \ottnt{M}  \ottsym{)}  \ottsym{)}$.  Although this function also takes the edit distance
  between $\ottnt{N}$ and $\ottnt{M}$ into account in the cost computation,
  $\textit{EditDist} \, \ottsym{(}  \ottnt{M}  \ottsym{,}  \textit{imitate} \, \ottsym{(}  \ottnt{N}  \ottsym{,}  \ottnt{M}  \ottsym{)}  \ottsym{)}$ does not count the distance between a
  hole in $\ottnt{N}$ and the corresponding subtree in $\ottnt{M}$, while
  $\textit{EditDist} \, \ottsym{(}  \ottnt{M}  \ottsym{,}  \ottnt{N}  \ottsym{)}$ in \COSTed{\ottnt{M}}{\ottnt{N}} counts the distance
  between a hole in $\ottnt{N}$ and the corresponding subtree in $\ottnt{M}$.
\end{itemize}

We call {\SYNTHESISbf} for the Procedure~\ref{program_synthesis}
instantiated with $\COSTbf{\ottnt{M}}{\ottnt{N}}$, {\SYNTHESISed} for one
instantiated with $\COSTed{\ottnt{M}}{\ottnt{N}}$, and {\SYNTHESISim} for one
instantiated with $\COSTim{\ottnt{M}}{\ottnt{N}}$.  Since {\SYNTHESISim}
treats the difference between a hole in a candidate $\ottnt{N}$ and the
corresponding subtree of guide term $\ottnt{M}$ as cost $0$,
{\SYNTHESISim} is expected to be more
aggressive in using the information of the guide term.  By comparing
{\SYNTHESISbf} against {\SYNTHESISed} and {\SYNTHESISim}, we can
discuss the merit of using neural machine translation with respect to
the brute-force search.


%


\begin{table}[t]
 \begin{center} \scriptsize

 \begin{tabular}{|c||c|c|c|c|c|c|c|c|c|c|c|c|} \hline
  Procedure & Sum & ED-0 & ED-1 & ED-2 & ED-3 & ED-4 & ED-5 & ED-6 & ED-7 & ED-8 & ED-10
  \\ \hline \hline
  $\SYNTHESISed_{D_\textit{tm}, 16}$
  & 1754.70 & 1.41 & 1.37 & 1.47 & 3.44 & 3.96 & 21.46 & 264.82 & 37.10 & 272.25 & N/A
  \\ \hline
  $\SYNTHESISim_{D_\textit{tm}, 16}$
  & 7546.57 & 1.46 & 1.80 & 1.91 & 5.53 & 5.21 & 29.99 & 1681.14 & 30.46 & 296.36 &N/A
  \\ \hline
  $\SYNTHESISed_{D_\textit{tm}, 32}$
  & 1336.19 & 1.49 & 1.51 & 11.52 & 1.98 & 8.10 & 35.17 & 59.14 & N/A & 243.59 & N/A
  \\ \hline
  $\SYNTHESISim_{D_\textit{tm}, 32}$
  & 2142.12 & 1.58 & 1.96 & 37.18 & 2.82 & 11.32 & 47.77 & 110.50 & N/A & 412.64 & N/A
  \\ \hline
  $\SYNTHESISed_{D_\textit{tm}, 64}$
  & 1173.34 & 1.38 & 1.69 & 1.95 & 3.41 & 8.86 & 30.35 & 40.37 & N/A & 425.25 & N/A
  \\ \hline
  $\SYNTHESISim_{D_\textit{tm}, 64}$
  & 3420.96 & 1.29 & 1.54 & 1.96 & 3.31 & 5.16 & 32.03 & 45.99 & N/A & 2688.28 & N/A
  \\ \hline
  $\SYNTHESISed_{D_{\beta\eta}, 16}$
  & 1587.47 & 1.44 & 1.61 & 2.10 & 3.19 & 21.52 & 3.57 & 81.98 & 247.83 & N/A & 1.96
  \\ \hline
  $\SYNTHESISim_{D_{\beta\eta}, 16}$
  & 2461.17 & 1.51 & 1.87 & 2.10 & 5.16 & 33.90 & 11.57 & 47.72 & 279.12 & N/A & 835.55
  \\ \hline
  $\SYNTHESISed_{D_{\beta\eta}, 32}$
  & 1308.47 & 1.49 & 1.86 & 3.08 & 4.15 & 1.74 & 13.95 & 102.53 & 3.42 & N/A & N/A
  \\ \hline
  $\SYNTHESISim_{D_{\beta\eta}, 32}$
  & 3316.54 & 1.41 & 1.90 & 3.57 & 4.13 & 1.94 & 17.52 & 299.16 & 10.43 & N/A & N/A
  \\ \hline
  $\SYNTHESISed_{D_{\beta\eta}, 64}$
  & 567.61 & 1.31 & 1.50 & 1.85 & 1.98 & 4.73 & 8.30 & 82.06 & N/A & 37.18 & N/A
  \\ \hline
  $\SYNTHESISim_{D_{\beta\eta}, 64}$
  & 640.44 & 1.20 & 1.55 & 2.04 & 2.16 & 6.97 & 10.02 & 90.30 & N/A & 38.67 & N/A
  \\ \hline
  \multicolumn{1}{|l||}{\SYNTHESISbf} & 28928.42+ & -- & -- & -- & -- & -- & -- & -- & -- & -- & -- \\ \hline
  \end{tabular}
 \end{center}

 \caption{Running time of the synthesis procedure (in seconds).  The
   column ``Procedure'' presents the procedure name with the used
   training set and the batch size.  The column ``Sum'' presents the
   the sum of running time for 100 test cases.  The column ED-$n$
   presents the average of running time for the test cases in which
   the edit distance between a guide term and the found proof term is
   $n$.  If a cell in the column ED-$n$ is marked N/A, it
   means that there was no test case in which the edit distance
   between a guide term and the found proof term was $n$.
   {\SYNTHESISbf} does not have data in the columns ED-$n$ since it
   ignores the guide term.}
 \label{tbl:eval-synthesis}
\end{table}

We generated a test dataset that consists of 100 types in the same way
as in \sect{exp-model} for evaluation.  We measured the running time
of each procedure with the models trained on different training
datasets and with different batch sizes.  The result is shown in
\tbl{eval-model}.  {\SYNTHESISbf} crashed due to a run-time memory
error in the 42nd test case; the value of the Sum column for
{\SYNTHESISbf} in \tbl{eval-model} reports the sum of the running
time until the 41st test case.


\paragraph{Discussion}

The two DNN-guided procedures {\SYNTHESISed} and {\SYNTHESISim} are
much faster than {\SYNTHESISbf}.  This indicates that guide terms
inferred by the trained models are indeed useful as a hint for proof search.

Comparing the synthesis procedures with models trained with different
datasets (i.e., $D_\textit{tm}$ and $D_{\beta\eta}$), we can observe
that the models trained with $D_{\beta\eta}$ often makes the synthesis
faster than the models trained on $D_\textit{tm}$.  This accords to
our expectation.  Although the result seems to be also largely
affected by the batch size used in the training phase, inspection
about the relation between the batch size and the performance of the
synthesis procedure is left as future work.

{\SYNTHESISim} is in general slower than {\SYNTHESISed} especially in
the cases where the edit distance is large.
We guess that this is due to the following reason.
{\SYNTHESISim} first explores a term that is almost the same as the
inferred guide term since \COSTim{\ottnt{M}}{-} calculates edit distances
by assuming that holes of proof candidates will be filled with
subterms of the guide term.
This strategy is risky because it wastes computation time if the
distance between the guide term and a correct proof is large.
%
The result in \tbl{eval-model} suggests that the current models tend
to infer a term such that it contains more errors if it is larger.
This inaccuracy leads to the worse performance of the current
implementation of {\SYNTHESISim}.  We think that {\SYNTHESISim}
becomes more efficient by improving the models.


%
%
The current implementation of {\SYNTHESISim} explicitly computes
$\textit{imitate} \, \ottsym{(}  \ottnt{N}  \ottsym{,}  \ottnt{M}  \ottsym{)}$ in the computation of the cost function.  This may
also affect the performance of {\SYNTHESISim}.  This could be improved
by optimizing the implementation of $\COSTim{\ottnt{M}}{\ottnt{N}}$.


To conclude the discussion, the guide by neural machine translation is
indeed effective in making the proof search more efficient.  We expect
that we can improve the performance by improving the accuracy of the
neural machine translation module.

\section{Related Work}
\label{sec:relatedwork}


Loos et al.~\cite{DBLP:journals/corr/LoosISK17} use a DNN to improve
the clause-selection engine of the Mizar
system~\cite{DBLP:conf/mkm/BancerekBGKMNPU15}.  In their work, the
input to the network is a set of unprocessed clauses and the negation
of a goal to be refuted; it is trained to recommend which unprocessed
clause to be processed.  They report that their architecture improves
the performance of the proof-search algorithm.  Our work shares the
same goal as theirs (i.e., DNN-assisted theorem proving) but tackles
in a different approach: they use a DNN for improving heuristics of an
automated theorem prover, whereas we use a DNN for translating a
proposition to its proof.  They observe that the combination of a DNN
and the conventional proof search is effective in expediting the overall
process, which parallels the design decisions of our proof synthesis, which uses
the proof suggested by a DNN as a guide for proof search.

As we mentioned in Section~\ref{sec:intro}, a proof-synthesis
procedure can be seen as a program-synthesis procedure via the
Curry--Howard isomorphism.  In this regard, the DNN-based program
synthesis~\cite{DBLP:journals/corr/DevlinUBSMK17,DBLP:journals/corr/BalogGBNT16}
are related to our work.  Devlin et
al.~\cite{DBLP:journals/corr/DevlinUBSMK17} propose an example-driven
program-synthesis method for string-manipulation problems.  They compare two
approaches to DNN-based program learning: \emph{neural synthesis},
which learns a program written in a DSL from input/output examples,
and \emph{neural induction}, which does not explicitly synthesize a
program but uses a learned model as a map for unknown inputs.  Balog et
al.~\cite{DBLP:journals/corr/BalogGBNT16} propose a program-synthesis method
for a functional DSL to manipulate integer lists.  Their
implementation synthesizes a program in two steps as we do: a DNN
generates a program from a set of input--output pairs; then, the
suggested program is modified for a correct program.  Both of Devlin et al.\ and
Balog et al.\ study
\emph{inductive} program synthesis that generates a program from given
input--output examples, while our work corresponds to program
synthesis from given specifications.
The state-of-the-art program synthesis with type
specifications~\cite{DBLP:conf/pldi/PolikarpovaKS16} is generating programs from
liquid types~\cite{DBLP:conf/cav/RondonBKJ12}, which allow for representing a
far richer specification than STLC.
We are currently investigating whether our proof-as-translation view is extended
to a richer type system (or, equivalently, a richer logic).

\section{Conclusion}
\label{sec:conclusion}

We proposed a proof-synthesis procedure for the intuitionistic
propositional logic based on neural machine translation.  Our
procedure generates a proof term from a proposition using the
sequence-to-sequence neural network.  The network is trained in
advance to translate the token-sequence representation of a
proposition to that of its proof term.  Although we did not carefully
tuned the network, it generates correct proofs to the almost half of
the benchmark problems.  We observed that an incorrect proof is often
close to a correct one in terms of the tree edit distance.  Based on
this observation, our procedure explores a correct proof using the
generated proof as a guide.  We empirically showed that our procedure
generates a proof more efficiently than a brute-force proof-search
procedure.  As far as we know, this is the first work that applies
neural machine translation to automated theorem proving.

As we mentioned in Section~\ref{sec:intro}, one of the main purposes
of the present paper is to measure the baseline of DNN-based automated
proof synthesis.  The result of our experiments in
Section~\ref{sec:exp} suggests that the current deep neural network
applied to automated proof synthesis can be trained so that it
generates a good guess to many problems, which is useful to make a
proof-search process efficient.

We believe that this result opens up several research directions that
are worth being investigated.  One of these directions is to extend
the target language.  Although we set our target to a small language
(i.e., the intuitionistic propositional logic) in the present paper,
automated proof synthesis for more expressive logic such as
Calculus of Construction~\cite{DBLP:journals/iandc/CoquandH88} and
separation logic~\cite{DBLP:conf/lics/Reynolds02,DBLP:conf/popl/IshtiaqO01} is
useful.  In
an expressive logic, we guess that we need more training data to avoid
overfitting.  To obtain such large amount of data, we consider using
an open-source repository of the proofs written in the language of
proof assistants such as Coq~\cite{Coq:manual} and
Agda~\cite{DBLP:conf/tldi/Norell09}.

Another future direction is to improve the performance of the neural
machine translation.  In general, the performance of a deep neural
network is known to be largely affected by how well the network is
tuned.  Besides the network itself, we guess that the performance
may be improved by tuning the representation of propositions and proofs.
For example, we used the infix notation to represent a proposition
(e.g., $\ottnt{S} \,  \mathord{\rightarrow}  \, \ottnt{T}$ for an implication), although a proof term for an
implication is an abstraction $\lambda  \mathit{x}  \ottsym{.}  \ottnt{M}$.  If we represent an
implication in the postfix notation (i.e., $(S, T){\rightarrow}$),
then the symbol ${\rightarrow}$ in the proposition and the symbol
$\lambda$ in the proof term comes closer in a training data, which may
lead to a better performance of sequence-to-sequence networks as is suggested by
Sutskever et al.~\cite{DBLP:journals/corr/SutskeverVL14}.

The current proof-search phase uses several variants of cost functions
to prioritize the candidates to be explored.  By tuning this function,
we expect that we can make the synthesis procedure faster.  We are
currently looking at the application of reinforcement
learning~\cite{Sutton:1998:IRL:551283} to automatically search for a
cost function that leads to a good performance of the overall
synthesis procedure.


\ifanon\else
\section*{Acknowledgments}
We would like to thank Takayuki Muranushi for making a prototype implementation
of the early proof synthesizer without DNNs.
We also appreciate Akihiro Yamamoto; the discussion with him leads to the
evaluation metrics used in this paper.
This paper is partially supported by JST PRESTO Grant Number
JPMJPR15E5, Japan.
\fi

\bibliographystyle{splncs03}
\bibliography{main}

\end{document}